\documentclass[twocolumn,10pt,prl, nofootinbib]{revtex4}

\usepackage[dvips]{graphicx}
\usepackage{epsfig,amsmath,amssymb,verbatim,mathrsfs,array,layout,textcomp,amssymb,latexsym,slashed,graphicx,hyperref,booktabs,color,cleveref}

\begin{document}
\title{Co-Decaying Dark Matter}

\author{Jeff Asaf Dror}\email{ajd268@cornell.edu}
\author{Eric Kuflik}\email{kuflik@cornell.edu}
\author{Wee Hao Ng}\email{wn68@cornell.edu}
\affiliation{Laboratory for Elementary Particle Physics, Cornell University,
Ithaca, NY 14850, USA}

\begin{abstract}
We propose a new mechanism for thermal dark matter freezeout, termed \emph{Co-Decaying Dark Matter}. Multi-component dark sectors with degenerate particles and out-of-equilibrium decays can co-decay to obtain the observed relic density. The dark matter density is exponentially depleted through the decay of nearly degenerate particles, rather than from Boltzmann suppression. 
The relic abundance is set by the dark matter annihilation cross-section, which is predicted to be boosted, and the decay rate of the dark sector particles. The mechanism is viable in a broad range of dark matter parameter space, with a robust prediction of an enhanced indirect detection signal. 
Finally, we present a simple model that realizes co-decaying dark matter. 
\end{abstract}

\maketitle

\section{Introduction}
The nature of dark matter (DM) is one of the most important open questions in  physics. The possibility that dark matter is a thermal relic with mass around the weak scale is intriguing, but has been under significant experimental pressure from direct detection~\cite{Cushman:2013zza,Akerib:2015rjg,Aprile:2015uzo} and at the LHC~\cite{Askew:2014kqa}. This motivates the study of models which are not constrained by these searches, but can still be discovered by indirect detection, where limits are weaker and have made rapid progress in recent years~\cite{Buckley:2013bha}.

Mechanisms for thermal dark matter freezeout usually rely on the DM remaining in chemical and thermal equilibrium with the Standard Model (SM) bath while non-relativistic, which leads to depletion of DM through Boltzmann suppression. 
 In this work we consider the possibility that part of the dark sector decays out of equilibrium with the SM. This delays the exponential suppression of the DM density well beyond the point where the DM candidate becomes non-relativistic. 

The mechanism, which we refer to as \emph{Co-Decaying Dark Matter}, has the following properties:
\begin{enumerate} 
\item The dark sector has decoupled from the SM before it becomes non-relativistic.
 \item The lightest dark sector particle decays into the SM out of equilibrium.
 \item The dark sector contains additional particles that are (approximately) degenerate with the decaying particle, and remain in chemical and thermal equilibrium with it until freezeout. One or more of these particles are DM candidates.  
\end{enumerate}
Co-decaying DM will be a generic feature of large dark sectors in which the lightest state decays. 
To illustrate the idea, we will focus on the simplified case of two degenerate dark sector particles: $A$ will be the DM candidate, and $B$ will be the decaying state, with sizable annihilations $AA \to BB$.  

\begin{figure}[t!]
  \begin{center} 
\includegraphics[width=.4 \textwidth]{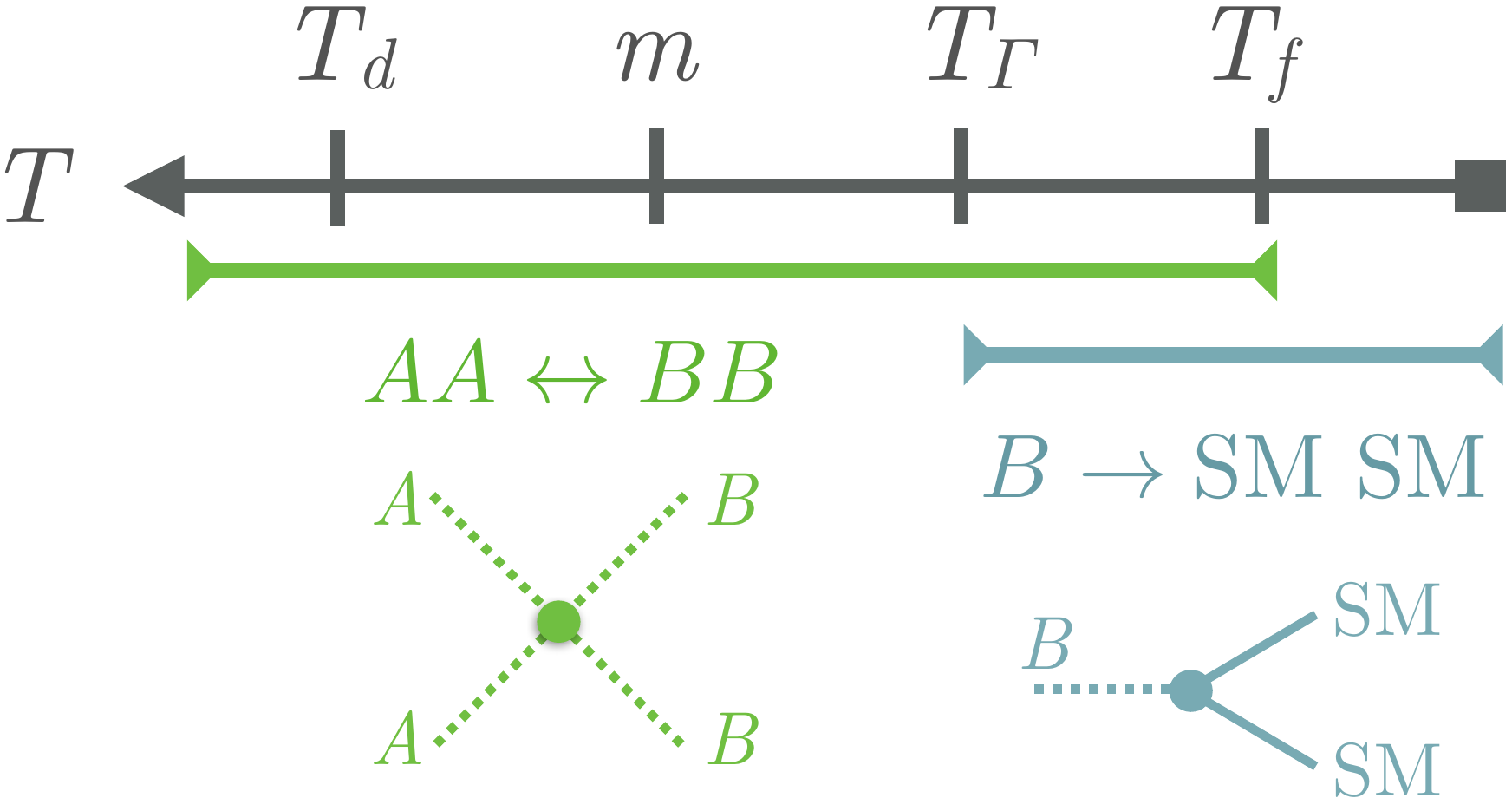} 
\end{center}
\caption{Co-decay dark matter timeline. At $ T _d  $ the SM and dark sector decouple; at $ T _\Gamma $ the decay of $ B$'s begin to deplete the dark sector density; and at $T _f $ the $ AA \leftrightarrow BB $ process freezes out, resulting in a relic abundance for the $ A $ particles. }
\label{fig:scematic}
\end{figure}

After the dark sector decouples from the SM bath, the $A$ and $B$ comoving entropy density is conserved, and their number density does not exponentially deplete when they become non-relativistic (in contrast to the  Weakly Interacting Massive Particle (WIMP)). Instead, the exponential suppression is delayed until the $B$'s begin decaying:
\begin{equation} 
n _A \sim n _B  \propto e ^{ - \Gamma_B t }
   \simeq e ^{ - \frac{1}{2} \Gamma_B/H }
 \,,
\end{equation} 
where $ n _{A,B} $ is the number density, $ \Gamma_B $ is the decay rate of the $ B $ particle, and $H$ is the Hubble parameter.  The $ A $ population tracks the $ B $ population until the $ AA \rightarrow  BB $ process cannot keep up with the expansion of the universe.
 At this point the $ A $ population freezes out and the $ B $'s continue to decay. The   relic density of $A$ is then set by both the annihilation rate, $\langle \sigma v \rangle$, as well as the $B$ decay rate, $\Gamma_B$.  A schematic illustration of the timeline for co-decaying DM is shown in Fig.~\ref{fig:scematic}.
 
 The delay in the starting point of exponential suppression from the temperature in which DM becomes non-relativistic to the temperature at which $B$-decay begins, causes freezeout to occur at later times than the WIMP. The DM relic density has less time to redshift to today, and therefore, must have a smaller density at freezeout.  In order to match the observed DM relic abundance a larger annihilation cross-section is required. This leads to a boosted indirect detection signal relative to WIMP models.
 
Previous work on multi-component dark sectors where interactions within the dark sector are necessary to get the correct dark matter relic abundance is extensive. Some examples including co-annihilating~\cite{Griest:1990kh,Baker:2015qna}, Secluded~\cite{Pospelov:2007mp}, SIMP~\cite{Hochberg:2014dra,Hochberg:2014kqa},  Cannibalizing~\cite{Pappadopulo:2016pkp,Kuflik:2015isi,Bernal:2015xba,Bernal:2015ova,Carlson:1992fn,Farina:2016llk} and Forbidden~\cite{Griest:1990kh,D'Agnolo:2015koa} DM. Additionally, models of particle decays affecting the relic abundance have been considered in~\cite{Feng:2003xh,Kaplinghat:2005sy,Farina:2015uea,Moroi:1999zb,Acharya:2009zt,Hall:2009bx,Berlin:2016vnh,Morrissey:2009ur,Cohen:2010kn,Bandyopadhyay:2011qm,Farina:2016llk}. The freezeout mechanism of co-decaying DM is unique, with differing phenomenology.  Furthermore, we emphasize that while we are mainly interested in the implications on dark matter, the dynamics studied here have a broad impact and can take place for any thermal relic.

In this Letter we study the co-decaying DM mechanism. We present an intuitive estimate of the relic density and check the results numerically using the Boltzmann equations. The  constraints and signals of co-decaying DM are described,  with a significant enhancement in the indirect detection signature. We conclude by presenting an explicit model realizing the phenomena.

\section{Freezeout and Relic Abundance}
\label{sec:crude}

The DM relic abundance can be solved in the standard sudden freezeout approximation, when $AA \to BB$ annihilations effectively stop:
\begin{equation}
n_{A, f} \langle \sigma v \rangle_f = H_f ~  \Longrightarrow  ~  \Omega_A   = \frac{s_0}{\rho_c }  \frac{\sqrt{g_{\star , m }}}{\sqrt{g_{\star , f }}}\frac{m H_m }{s_m } \frac{x_f}{ \langle  \sigma v \rangle_f}.
\label{suddenf}
\end{equation}
Here $ m  $ is the DM mass, $x_i=m/T_i$, 
$s$ is the entropy density of the  SM bath,
and the subscripts $m$ and $f$ denote quantities at temperatures $T=m$ and freezeout, respectively\footnote{Throughout this section we will neglect the differences in effective entropy degrees of freedom $g_{\star s}$ and effective energy degrees of freedom $g_{\star}$.}. Note that Eq.~(\ref{suddenf}) is identical to the standard WIMP scenario. However, for co-decaying DM,  we will see that $x_f \gg 1$, leading to a boosted annihilation cross section relative to the standard WIMP case, where $x_f $ $ \simeq 20$. 

We now compute the SM and dark sector temperatures at freezeout. To this end, we study the temperature evolution of the dark sector through the three stages depicted in Fig.~\ref{fig:scematic}: from the time of decoupling of the dark sector from the SM ($ T _d $), to the onset of the $ B $ decay ($T _\Gamma $), and until freezeout of the $ AA \rightarrow BB $ annihilations ($ T _f $). We use the $ d $, $ \Gamma $, and $ f $ subscripts throughout to denote quantities evaluated at these stages, respectively, and primes to denote dark sector (total $A+B$) quantities. 

At high temperatures, $A$ and $B$ decouple from the SM plasma when relativistic.  The entropy densities in each sector are separately conserved until the decay of $B$ begins, and therefore
\begin{equation}
s^\prime_\Gamma = \frac{s^\prime_d}{s_d}s_\Gamma  \equiv \xi s_\Gamma \,,
\label{xi}
\end{equation}
The dark sector number density at the onset of decay, roughly when ${\Gamma_B \simeq H_\Gamma}$, is given by the second law of thermodynamics for non-relativistic particles: 
\begin{equation}
n_\Gamma^\prime = \frac{T_\Gamma^\prime} {m-\mu^\prime_\Gamma+ \frac{5}{2} T_\Gamma^\prime }\xi s_\Gamma\,,
\end{equation} 
where $ \mu ' $ is the chemical potential of $ A $ and $B$.

While the $AA \leftrightarrow BB$ process is fast, the $A$ density matches the $B$ density. Taking the number of degrees of freedom in $A$ and $B$ to be equal (which we will assume throughout the paper for simplicity), the total dark sector density at the time of $AA \leftrightarrow BB$ freezeout is
\begin{equation}
n_f^\prime a_f^3 =  n_\Gamma^\prime  a_\Gamma^3 e^{-\frac{1}{2}\Gamma_B (t_f -t_\Gamma)}  \simeq \frac{ \xi s_\Gamma a_\Gamma^3}{x_\Gamma^\prime-\frac{\mu^\prime}{T_\Gamma^\prime}+ \frac{5}{2} } e^{-\frac{1}{4} \frac{\Gamma_B}{H_f}}\,.
\label{numf}
\end{equation} 
where $a$ is the cosmic scale factor. The $A$ abundance is hence depleted through the decay of $B$ particles. Using Eq.~(\ref{suddenf}) with Eq.~(\ref{numf}), the temperature at freezeout is given by
\begin{equation}
x_f \simeq \frac{2  }{\sqrt{\Gamma_B / H_m} }\log^{1/2} \frac{\frac{2}{ \sqrt{ \pi }}\frac{s_m}{H_m} \xi  \sigma  }{ x _f \sqrt{ x_f^{\prime }} x_\Gamma^\prime (1-\frac{\mu^\prime_\Gamma}{m} + \frac{5}{2x_\Gamma^\prime} ) } \,,
\label{xf}
\end{equation}
where  $  n _{A, f } = \frac{1}{2}  n _f ' $ and for brevity we have dropped ratios of $ g _{ \star } $. 
Here we have taken
\begin{equation}
\langle \sigma v \rangle =  \frac{4}{\sqrt{\pi}}\frac{\sigma}{ \sqrt{x^{\prime}}}
\label{lk}
\end{equation}
for $x^\prime\gg 1$ and $s$-wave scattering, where $\sigma$ is  the $2 \to 2$ cross-section at threshold.  (For reference,  note that  the observed relic density for a WIMP would require $\sigma   \simeq 10^{-36}~ {\rm cm}^2$.) 
 Since $\Gamma_B / H_m$ may be as small as $10^{-18}$ (see Fig.~\ref{fig:validity}), $x_f$ may be as large as $10^8$. 

 \begin{figure}[t!]
  \begin{center} 
\includegraphics[width=.47\textwidth]{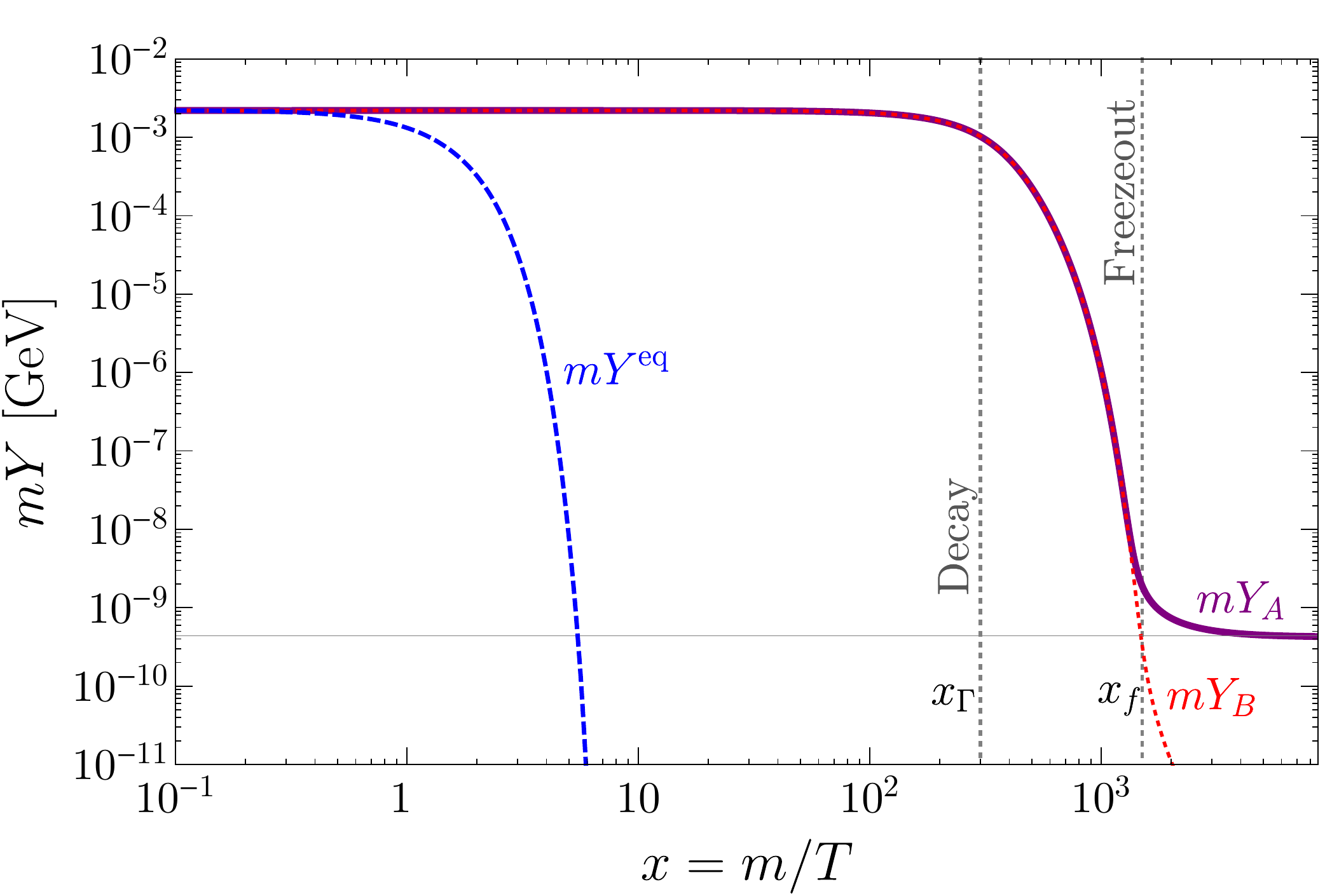} 
\end{center}
\caption{Yields ($ Y \equiv n  / s $) as a function of SM temperature without cannibalism for a benchmark point $g_A=g_B=1$, $ m = 1 \text{ GeV} $, $ \sigma  = 1 \times 10 ^{ -30} {\rm~cm}^2 $, $\Gamma_B=6\times 10^{-23}~\rm GeV$. The (\textbf{{\color[rgb]{.5,0,.5}purple}}/solid) and (\textbf{{\color[rgb]{1,0,0}red}}/dotted) lines show the yield for $A$ and $B$ particles, respectively.  For comparison, the (\textbf{{\color[rgb]{0,0,1}blue}}/dashed) line shows the yield assuming the DM was in chemical and thermal equilibrium.  For this choice of parameters $x_\Gamma \simeq 300 $, while freezeout occurs at $x_f \simeq 1500$. The dark temperature at freezeout is $x_f^\prime \simeq 5\times 10^6$.
}
\label{fig:Boltz}
\end{figure}

 The chemical potential and dark temperature will depend on whether number changing processes are active in the $A,B$ system, e.g., $3 \to 2$ processes. Without number changing processes, the comoving entropy and number densities are separately conserved in the dark sector between decoupling and decay ($s'_\Gamma/s_\Gamma = s'_d/s_d$ and $n'_\Gamma/s_\Gamma = n'_d/s_d$). This  decreases the dark temperature relative to the SM temperature, while inducing a chemical potential:
\begin{equation}
x_{\Gamma}^\prime \simeq\frac{1}{3.7} \left(  \frac{g_{\star  , d}}{g_{\star ,\Gamma}} \right)^{\frac{2}{3}} x_{\Gamma} ^2 , ~~\frac{\mu^\prime_\Gamma}{m} \simeq  1 - \frac{ 3 }{ 2 x' _\Gamma } ~~\rm  (w/o~canb),
\label{darkTnocann}
\end{equation}

In contrast, if number changing processes are active, cannibalization can occur~\cite{Carlson:1992fn}.  The SM temperature decreases exponentially relative to the dark sector, while the chemical potential is held fixed ($\mu^\prime =0$). Using conservation of comoving entropy in the hidden sector, one finds
\begin{equation}
x_{\Gamma}^\prime \simeq   \log \frac{ x_{\Gamma}^3 }{3.\,\xi \,x_{\Gamma}^{\prime\,1/2} g_{\star,{\Gamma}} },  ~~\frac{\mu^\prime_\Gamma}{m} =0~~\rm  (w~canb).\label{darkTcann}
\end{equation}

In both cases, the dark temperature at freezeout is redshifted from the temperature at decay,
\begin{equation} 
x_f^\prime \simeq x_\Gamma^\prime \left(\frac{a_f}{a_\Gamma}\right)^2 \sim x_\Gamma^\prime \left(\frac{x_f}{x_\Gamma}\right)^2 \label{Tfdark}\,.
\end{equation} 
Note that the dark matter will have a large energy density before it decays,
and may come to dominate the energy density of the universe. When the DM decays, it will release a significant amount of entropy and reheat the SM bath. However, since the reheating occurs before DM freezeout, the entropy dump does not dilute the DM relic abundance. 
The most important effects are a delay in the start of the decay and a modification to the final relationship in Eq.~(\ref{Tfdark}). These effects are taken into account in the numerical solutions to the Boltzmann equations and in computing the viable parameter space.

Combining Eqs. \labelcref{suddenf,xf,lk,darkTnocann,darkTcann,Tfdark}, 
the relic abundance in the absence of cannibalization and when cannibalization is active throughout is:
\begin{equation}    
   \dfrac{\Omega_A}{\Omega_{\rm DM}}   \!\simeq\! \left(\dfrac{10^{-36} }{\sigma/\rm cm^{2}} \right)   \times \left\{
\begin{array}{lr}
\!\left(\dfrac{m}{\rm GeV}\right)   
  \left( \dfrac{ 10^{-18}}{\Gamma_B/m} \right)
       &    \rm (w/o~canb),\\
\!\!  \left(\dfrac{m}{\rm GeV}\right)^{\!\frac{1}{2}}
 \left(\dfrac{10^{-17}}{\Gamma_B/m} \right)^{\!\frac{1}{2}}
      &    \rm  (w~canb).
     \end{array}
   \right.
\label{relicdensity2}
\end{equation}
where we have taken, $g_{\star , \, d}=106.75$, and $\Omega_{\rm DM}=0.27$~\cite{Ade:2015xua}. Here and throughout we will take the entropy density ratio at decoupling, defined in Eq.~(\ref{xi}), to be $\xi={(g_A+g_B)} /g_{\star, \, d}\simeq 0.02$.

Generically in any given model, one expects number changing self-interactions to be present, which leads to  some amount of cannibalization. Additionally, in much of parameter space cannibalization can shut off before decays begin. Therefore, a realistic scenario will likely be between the two limiting cases in Eq.~(\ref{relicdensity2}).

 \section{Boltzmann equations}\label{sec:boltz}

 We now present a numerical study of co-decaying dark matter. To track the number densities of $ A $ and $ B $ as well as the dark temperature $T^\prime$,  $ 3 $ different equations are required:
\begin{equation}
\begin{array}{l}
\dot{n}_A + 3 H n_A=- \langle \sigma v \rangle (n_A^2 - n_B^2) \,,\\ 
\dot{n}_{A+B} + 3 H n_{A+B} =  - \left(  \langle \Gamma_B \rangle_{T^\prime} n_B -  \langle \Gamma_B \rangle_T n^{{\rm eq}}_T\right)   \,, \\
\dot{\rho}_{A+B}+ 3 H ({\rho}_{A+B}+ P_{A+B}) = -  m \Gamma_B \left(n_B - n^{{\rm eq}}_T \right)  \,,
\end{array}
\label{BE}
\end{equation}
where $\langle \Gamma_B \rangle_{T(T^\prime)} /\Gamma_B = m \langle E_B^{-1} \rangle_{T(T^\prime)}$ is the thermally averaged inverse boost factor over the DM (SM) phase-space distributions.
Time derivatives can be related to derivatives of the SM temperature $T$ using the Friedman equation and second law of thermodynamics,
\begin{eqnarray} 
H^2 &\equiv& \left(\frac{\dot{a}}{a}\right)^2 = \frac{8 \pi  G}{3} (\rho +\rho^\prime) \,,\\ 
a\frac{d}{da} ( s \,  a^3) & =  &  \frac{ 1 }{ T} \frac{ d ( \rho \, a ^4 ) }{ d a } =  \frac{m \Gamma_B }{H T} \left(n_B - n^{{\rm eq}}_T \right) a^3\,, \nonumber
\end{eqnarray} 
where $G$ is the gravitational constant. 

If number-changing processes in the dark sector are present, such as $3 \to 2$ processes, then there are additional terms in the number density equations of the form
\begin{equation} 
 -  \left\langle \sigma v ^2 \right\rangle _{ ijk \rightarrow lm}  (n_i n_j n_k - n_l n_m n^{\rm eq} ) \,,
\end{equation} 
where $n_i$ can be $n_A$ or $n_B$. 

The Boltzmann equations, Eq. (\ref{BE}), are straightforward to solve numerically, and the results for a benchmark point are given in Fig.~\ref{fig:Boltz}. As shown, the dark sector does not follow the equilibrium distribution; instead it undergoes exponential decay at a later time. When the dark matter becomes non-relativistic ($x\simeq 1$), the co-moving number density remains constant, until decay  begins ($x\simeq x_\Gamma$).  The $A$ density matches the $B$ density until freezeout ($x\simeq x_f$) , where the DM candidate $ A $ decouples, while $ B $ continues to decay. For smaller $\Gamma_B$, the co-moving number density remains constant for longer and the decays will begin later. Depending on the size of $\Gamma_B$, the cross section needed to decouple at the correct time, and match the observed relic abundance, can be orders of magnitude larger than those of the WIMP scenario. The solutions to the Boltzmann equations match well the analytic estimates given by Eq.~(\ref{relicdensity2}).

\section{Signatures and constraints}
\begin{figure*}[th!]
  \begin{center} 
\includegraphics[width= .49\textwidth]{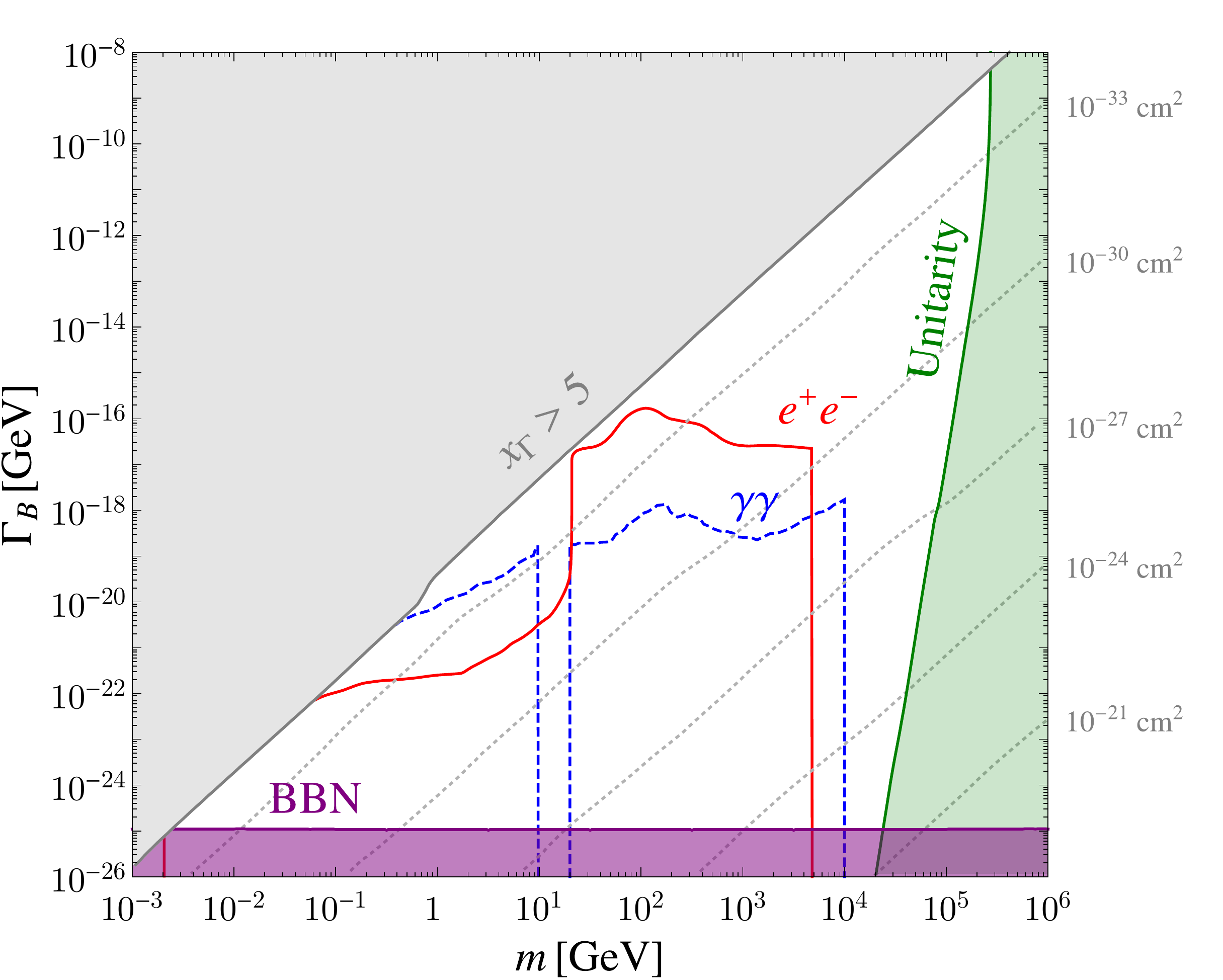} \hfill
\includegraphics[width= .49\textwidth]{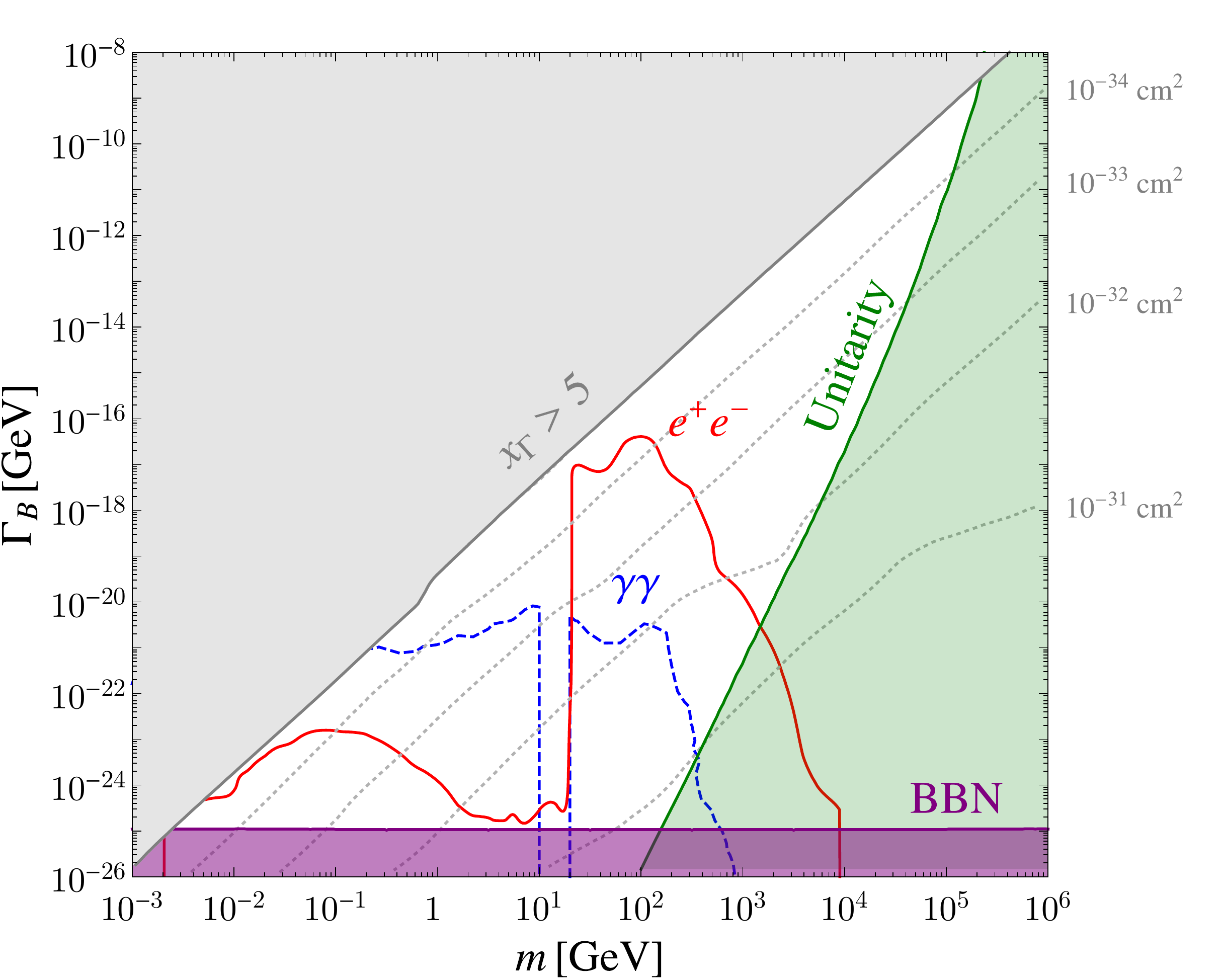} 
\end{center}
\caption{The viable parameter space for co-decaying dark matter assuming no cannibalization ({\bf Left}), and a cannibalizing dark sector  ({\bf Right}).
 The central white region shows the range of validity of the model. The different regions show constraints from $N_{\rm eff}$ ({\bf \color[rgb]{.5,0,.5}purple}); DM decays out of equilibrium ({\bf \color[rgb]{0.3,0.3,0.3}gray}); unitarity constraints ({\bf \color[rgb]{0,0.5,0}green}); and indirect detection assuming decays into $ e ^+ e ^- $ ({\bf \color[rgb]{1,0,0}red}/solid) or $ \gamma \gamma $ ({\bf \color[rgb]{0.,0,1}blue}/dashed), excluding the region below the curve. The gap in the $ \gamma \gamma $ limit between $ 10 - 20 \text{ GeV} $ is due to thresholds used in the two recasts. The {\bf \color[rgb]{0.7,0.7,0.7}light gray} dotted lines represent contours of constant $ \sigma $ with values indicated on the right.\label{fig:validity}}
\end{figure*}

We now discuss the signatures and constraints of co-decaying dark matter , whose parameter space is characterized by $ m $, $ \Gamma_B $, and $  \sigma $. The viable parameter space is summarized in Fig.~\ref{fig:validity}, where the dotted gray lines represent contours of constant $ \sigma $. As expected by the rough estimate in Eq.~\ref{relicdensity2}, the cross section contours are much more widely spaced without cannibalization than without.

First, the co-decay setup requires $B$ to decay out of equilibrium; otherwise the dark matter candidate will be Boltzmann suppressed when it becomes non-relativistic, effectively reducing to the WIMP scenario. This corresponds to $ x _\Gamma \gtrsim 1 $, though requiring that the DM does not re-thermalize with the SM imposes $ x _\Gamma \gtrsim 5 $. This is depicted by the gray shaded area in Fig.~\ref{fig:validity}.

Next we consider constraints on $ N _{ \text{eff}} $~\cite{DiValentino:2016ikp,Boehm:2013jpa}. This gives the rough condition that the DM decays before big bang nucleosynthesis (BBN), $ \Gamma_B \gtrsim H _{m_e} $. This is depicted by the shaded purple regions in Fig.~\ref{fig:validity}.

Unitarity places constraints on the size of the thermally averaged cross section. The requirement of unitarity is given for $ s $-wave scattering by~\cite{Griest:1989wd},
\begin{equation} 
  \left\langle \sigma v \right\rangle _{f} \le \frac{ 4\pi \left\langle v ^{-1} \right\rangle _{ f} }{ m ^2 }  \quad  \Longrightarrow  \quad \sigma   \lesssim \frac{\pi \sqrt{ 2 }}{ m ^2 } x _f ^{ \prime }\,,
\end{equation} 
where $ \left\langle v ^{-1} \right\rangle  _{f}\simeq  \sqrt{ 2 x ' _f  / \pi } $ is the thermally averaged inverse velocity. 
The severity of the bound is dependent on whether or not the dark sector is cannibalizing. Without cannibalization, fixing the relic density corresponds to $ \sigma   \propto 1/ \Gamma_B $ and $ x _f ' \propto  1 / \Gamma_B $, and thus the unitarity bound is roughly $ \Gamma_B $-independent. On the other hand, with cannibalization, $ \sigma   \propto 1/ \sqrt{ \Gamma_B } $ and $ x _f ' $  is only log-dependent on $ \Gamma_B  $, and the unitarity bound reads:
\begin{equation}     
   \frac{m }{  \text{ 100 TeV} } \lesssim \left\{
\begin{array}{ll}
 1  & \text{(w/o canb)}, \\ 
100\left(\frac{\Gamma_B}{\rm GeV}\right) ^{ 1/6}  & \text{(w/ canb)}\,.
\end{array}   \right.
\end{equation} 
These relations are modified for small $ \Gamma $ by matter-domination effects. The resulting unitarity bounds are shown in the green shaded regions in Fig.~\ref{fig:validity}.

Since co-decaying DM is decoupled from the SM, it is difficult to discover using direct detection or direct production. The signature from indirect detection is, however, enhanced with respect to WIMP candidates due to the large thermally averaged cross section. This makes indirect detection a powerful tool to probe co-decaying DM.

We map the current constraints from telescope and satellite data on the $ ( m, \Gamma_B ) $ parameter space, using the analyses of Refs.~\cite{Essig:2013goa,Elor:2015bho}.
 The constraint on our four-body final state from two-body final states analyzed in~\cite{Essig:2013goa,Elor:2015bho} are obtained by rescaling the mass and cross-section limits appropriately. 
For illustration, we plot the full constraints from $ B $ decays  into only $ e ^+ e ^- $ (red, solid) or into only $ \gamma \gamma $ (blue, dashed) in  Fig.~\ref{fig:validity}, excluding the region below the curves. 

 Lastly we note that co-decaying dark matter is not constrained by the Cosmic Microwave Background, since the thermally averaged cross-sections is always velocity suppressed.

The combined allowed parameter space is shown in Fig.~\ref{fig:validity}, without cannibalization (left-panel) and with cannibalization (right-panel). We learn that  co-decaying dark matter can occur over a broad range of DM masses, spanning an MeV up to hundreds of TeV, and decay rates spanning many orders of magnitude.

\label{sec:sig}
\section{Mass splitting}

Thus far, we focused on degenerate dark sector particles, which can result from an underlying symmetry. However, a realistic model may include symmetry-breaking effects, which can lift the degeneracy. It is then important to understand the effect of mass-splittings on the co-decaying DM framework. We leave a detailed study of the phenomenology of co-decays with mass splittings to future work~\cite{future} and highlight the expected features here. 

If $ m _A > m _B $, the  co-decay mechanism remains conceptually unchanged. However, for mass splitting  $\mathcal{O}( \%)$ or more, the  parameter space to produce the observed relic abundance can differ significantly. To understand this, consider s-wave annihilation, which can proceed as zero temperature in the presence of mass-splittings.  Comparing the annihilation rates at large $x^\prime$, we have  
\begin{equation} 
\langle \sigma v \rangle_{m_A > m_B} \simeq \frac{\sqrt{\pi x^\prime}}{2} \langle \sigma v \rangle_{m_A = m_B}
\end{equation} 
 for fixed matrix-element. Since freezeout occurs for $x^\prime \gg 1$, obtaining the observed relic abundance requires $ \sigma  $ smaller than in the degenerate case. 
 
If $ m _A < m _B $,  then annihilations proceed off the exponential tail of $A$'s velocity distribution,
$
\left\langle \sigma v \right\rangle _{ AA \rightarrow BB } \propto e ^{ - 2 \Delta x '},$,
where $ \Delta \equiv ( m _B - m _A ) / m _A $. This exponential suppression of the cross-section significantly alters the parameters required to produce the correct relic density. 
 
\section{Model}
\label{sec:model}

Having described the general framework,  we now present a simple model where co-decay can drive dark matter freezeout. Consider a dark SU(2)$_D$ gauge theory with coupling $g_D$, and a dark Higgs doublet $\Phi_D$,
\begin{equation}
\mathcal{L} \supset D^\mu \Phi_D^\dagger D_\mu \Phi_D -\frac{1}{4} F_D^{a,\mu \nu} F^a_{D,\mu \nu} - \lambda_D \left(\Phi_D^\dagger \Phi_D - \frac{v_D^2}{2} \right)^2\,,
\end{equation}
The dark Higgs' VEV, $ v _D / \sqrt{2} $,  spontaneously  breaks SU(2)$_D$. All three dark gauge bosons acquire masses $m_D = \frac{1}{2} g_D v_D$, while the dark Higgs boson, $h _D $, gains a mass $m_{ h _D } = \sqrt{2\lambda_D} v_D$. The stability and degeneracy of the gauge bosons are ensured by an unbroken SU(2) custodial symmetry. We take $m_{ h _D } \gg m_D$, which decouples the dark Higgs.

We introduce a dimension-six operator, which explicitly breaks the custodial symmetry down to U(1),
\begin{equation}
\mathcal{L} \supset \frac{(\Phi_D^\dagger D^\mu \Phi_D)(\Phi^\dagger D_\mu \Phi)}{\Lambda^2}\,,
\end{equation}
where $\Phi$ is the SM Higgs doublet. This can be generated by integrating out heavy fermions charged under both SU(2)$_D$ and the SM gauge symmetry, SU(2)$_L$. This operator  mixes the gauge boson $Z_D \equiv W_D^3$ and the $Z$ boson,  decaying $Z_D$ to the SM.
 The remaining gauge bosons $W_D^\pm \equiv (W_D^1 \mp iW_D^2)/\sqrt{2}$ are stable since they are the lightest particles charged under the unbroken U(1) custodial symmetry. 

The $W^\pm_D$ are stable and play the role of $A$, while the nearly-degenerate $Z_D$ plays the role of $B$. For $ m_D \sim  \text{ GeV} $ and $\Lambda \sim 10$'s TeV,  negligible mass differences between $W_D^\pm$ and $Z_D$ are generated, and corrections to electroweak precision observables are small. Number-changing processes, e.g.,  $Z_D Z_D Z_D \to W_D^+ W_D^-$, are large and cannibalization effects must be taken into account.

This model can be mapped onto the constraints of the previous sections using
\begin{equation} 
\sigma = \frac{688 }{3} \frac{\alpha_D^2}{m_D^2}\,, ~~~~~~~ \Gamma_{Z_D} = \frac{1}{48 \pi^2 \alpha_D^2} \frac{m_D^5}{\Lambda^4} |g|^2\,,
\end{equation} 
where $|g|^2  \equiv \sum_i |g_i|^2 \left(|g_V^i|^2 +|g_A^i|^2 \right) $, $g_V~(g_A)$ is the vector (axial) coupling of the fermion $i$ to the $Z$-boson.

Lastly, we comment on further model building directions. To build a viable model one needs a approximate symmetry to achieve degeracy between the lightest dark states, but whose breaking induces a decay into the SM. In this section we considered the possibility that a remnant of a broken SU(2) gauge symmetry protects the masses, however interesting alternatives include flavor symmetries or supersymmetry, both of which could play a role in a larger framework. Depending on the type of symmetry used to ensure the degenaracy, this may or may not induce significant cannibalization.  
\begin{acknowledgments}
{\em Acknowledgments ---} We thank Marco Farina for collaboration in early stages of the project. We are grateful to Spencer Chang, Tim Cohen, Graham Kribs, Michelle Papucci, Tracy Slatyer, and Hitoshi Murayama for useful discussions. We especially thank Yonit Hochberg for useful discussions and comments on the manuscript. This work was supported in part by the NSF through grant PHY-1316222. JD is supported in part by the NSERC Grant PGSD3-438393-2013. EK is supported by a Hans Bethe Postdoctoral Fellowship at Cornell. 
\end{acknowledgments}

Note added: During the preparation of this work we became aware of Ref.~\cite{Farina:2016llk} which considers a similar scenario.
\bibliographystyle{h-physrev}
\bibliography{Codecay}

\begin{thebibliography}{10}

\bibitem{Cushman:2013zza}
P.~Cushman {\em et~al.},
\newblock {Working Group Report: WIMP Dark Matter Direct Detection},
\newblock in {\em {Proceedings, 2013 Community Summer Study on the Future of
  U.S. Particle Physics: Snowmass on the Mississippi (CSS2013): Minneapolis,
  MN, USA, July 29-August 6, 2013}}, 2013, 1310.8327.

\bibitem{Akerib:2015rjg}
LUX, D.~S. Akerib {\em et~al.},
\newblock Phys. Rev. Lett. {\bf 116}, 161301 (2016), 1512.03506.

\bibitem{Aprile:2015uzo}
XENON, E.~Aprile {\em et~al.},
\newblock JCAP {\bf 1604}, 027 (2016), 1512.07501.

\bibitem{Askew:2014kqa}
A.~Askew, S.~Chauhan, B.~Penning, W.~Shepherd, and M.~Tripathi,
\newblock Int. J. Mod. Phys. {\bf A29}, 1430041 (2014), 1406.5662.

\bibitem{Buckley:2013bha}
J.~Buckley {\em et~al.},
\newblock {Working Group Report: WIMP Dark Matter Indirect Detection},
\newblock in {\em {Proceedings, Community Summer Study 2013: Snowmass on the
  Mississippi (CSS2013): Minneapolis, MN, USA, July 29-August 6, 2013}}, 2013,
  1310.7040.

\bibitem{Griest:1990kh}
K.~Griest and D.~Seckel,
\newblock Phys. Rev. {\bf D43}, 3191 (1991).

\bibitem{Baker:2015qna}
M.~J. Baker {\em et~al.},
\newblock JHEP {\bf 12}, 120 (2015), 1510.03434.

\bibitem{Pospelov:2007mp}
M.~Pospelov, A.~Ritz, and M.~B. Voloshin,
\newblock Phys. Lett. {\bf B662}, 53 (2008), 0711.4866.

\bibitem{Hochberg:2014dra}
Y.~Hochberg, E.~Kuflik, T.~Volansky, and J.~G. Wacker,
\newblock Phys. Rev. Lett. {\bf 113}, 171301 (2014), 1402.5143.

\bibitem{Hochberg:2014kqa}
Y.~Hochberg, E.~Kuflik, H.~Murayama, T.~Volansky, and J.~G. Wacker,
\newblock Phys. Rev. Lett. {\bf 115}, 021301 (2015), 1411.3727.

\bibitem{Pappadopulo:2016pkp}
D.~Pappadopulo, J.~T. Ruderman, and G.~Trevisan,
\newblock Phys. Rev. {\bf D94}, 035005 (2016), 1602.04219.

\bibitem{Kuflik:2015isi}
E.~Kuflik, M.~Perelstein, N.~R.-L. Lorier, and Y.-D. Tsai,
\newblock Phys. Rev. Lett. {\bf 116}, 221302 (2016), 1512.04545.

\bibitem{Bernal:2015xba}
N.~Bernal and X.~Chu,
\newblock JCAP {\bf 1601}, 006 (2016), 1510.08527.

\bibitem{Bernal:2015ova}
N.~Bernal, X.~Chu, C.~Garcia-Cely, T.~Hambye, and B.~Zaldivar,
\newblock JCAP {\bf 1603}, 018 (2016), 1510.08063.

\bibitem{Carlson:1992fn}
E.~D. Carlson, M.~E. Machacek, and L.~J. Hall,
\newblock Astrophys. J. {\bf 398}, 43 (1992).

\bibitem{Farina:2016llk}
M.~Farina, D.~Pappadopulo, J.~T. Ruderman, and G.~Trevisan,
\newblock (2016), 1607.03108.

\bibitem{D'Agnolo:2015koa}
R.~T. D'Agnolo and J.~T. Ruderman,
\newblock Phys. Rev. Lett. {\bf 115}, 061301 (2015), 1505.07107.

\bibitem{Feng:2003xh}
J.~L. Feng, A.~Rajaraman, and F.~Takayama,
\newblock Phys. Rev. Lett. {\bf 91}, 011302 (2003), hep-ph/0302215.

\bibitem{Kaplinghat:2005sy}
M.~Kaplinghat,
\newblock Phys. Rev. {\bf D72}, 063510 (2005), astro-ph/0507300.

\bibitem{Farina:2015uea}
M.~Farina,
\newblock JCAP {\bf 1511}, 017 (2015), 1506.03520.

\bibitem{Moroi:1999zb}
T.~Moroi and L.~Randall,
\newblock Nucl. Phys. {\bf B570}, 455 (2000), hep-ph/9906527.

\bibitem{Acharya:2009zt}
B.~S. Acharya, G.~Kane, S.~Watson, and P.~Kumar,
\newblock Phys. Rev. {\bf D80}, 083529 (2009), 0908.2430.

\bibitem{Hall:2009bx}
L.~J. Hall, K.~Jedamzik, J.~March-Russell, and S.~M. West,
\newblock JHEP {\bf 03}, 080 (2010), 0911.1120.

\bibitem{Berlin:2016vnh}
A.~Berlin, D.~Hooper, and G.~Krnjaic,
\newblock Phys. Lett. {\bf B760}, 106 (2016), 1602.08490.

\bibitem{Morrissey:2009ur}
D.~E. Morrissey, D.~Poland, and K.~M. Zurek,
\newblock JHEP {\bf 07}, 050 (2009), 0904.2567.

\bibitem{Cohen:2010kn}
T.~Cohen, D.~J. Phalen, A.~Pierce, and K.~M. Zurek,
\newblock Phys. Rev. {\bf D82}, 056001 (2010), 1005.1655.

\bibitem{Bandyopadhyay:2011qm}
P.~Bandyopadhyay, E.~J. Chun, and J.-C. Park,
\newblock JHEP {\bf 06}, 129 (2011), 1105.1652.

\bibitem{Ade:2015xua}
Planck, P.~A.~R. Ade {\em et~al.},
\newblock Astron. Astrophys. {\bf 594}, A13 (2016), 1502.01589.

\bibitem{DiValentino:2016ikp}
E.~Di~Valentino, S.~Gariazzo, M.~Gerbino, E.~Giusarma, and O.~Mena,
\newblock Phys. Rev. {\bf D93}, 083523 (2016), 1601.07557.

\bibitem{Boehm:2013jpa}
C.~Boehm, M.~J. Dolan, and C.~McCabe,
\newblock JCAP {\bf 1308}, 041 (2013), 1303.6270.

\bibitem{Griest:1989wd}
K.~Griest and M.~Kamionkowski,
\newblock Phys. Rev. Lett. {\bf 64}, 615 (1990).

\bibitem{Essig:2013goa}
R.~Essig, E.~Kuflik, S.~D. McDermott, T.~Volansky, and K.~M. Zurek,
\newblock JHEP {\bf 11}, 193 (2013), 1309.4091.

\bibitem{Elor:2015bho}
G.~Elor, N.~L. Rodd, T.~R. Slatyer, and W.~Xue,
\newblock JCAP {\bf 1606}, 024 (2016), 1511.08787.

\bibitem{future}
J.~A. Dror, E.~Kuflik, and W.~H. Ng,
\newblock (2016),
\newblock {to appear}.

\end{thebibliography}

\end{document}